\documentclass{article}
\usepackage{spconf,amsmath,graphicx,hyperref}
\usepackage{lineno}
\usepackage{soul,xcolor}
\usepackage{multicol}
\usepackage{multirow}
\usepackage{enumitem}
\usepackage{adjustbox}
\usepackage{subcaption}
\usepackage{bm, booktabs}
\usepackage{xcolor}

\title{Test Time Adaptation for Speech Emotion Recognition}

\name{Jiaheng Dong$^{*\dagger}$, Hong Jia$^{*\ddagger}$\thanks{*Equal Contribution}, Ting Dang$^{\dagger}$}
\address{
$^{\dagger}$The University of Melbourne, Australia \
$^{\ddagger}$The University of Auckland, New Zealand \
}
\begin{document}
\ninept
\maketitle

\begin{abstract}
The practical utility of Speech Emotion Recognition (SER) systems is undermined by their fragility to domain shifts, such as speaker variability, the distinction between acted and naturalistic emotions, and cross-corpus variations. While domain adaptation and fine-tuning are widely studied, they require either source data or labeled target data, which are often unavailable or raise privacy concerns in SER. Test-time adaptation (TTA) bridges this gap by adapting models at inference using only unlabeled target data. Yet, having been predominantly designed for image classification and speech recognition, the efficacy of TTA for mitigating the unique domain shifts in SER has not been investigated. In this paper, we present the first systematic evaluation and comparison covering 11 TTA methods across three representative SER tasks. The results indicate that backpropagation-free TTA methods are the most promising. Conversely, entropy minimization and pseudo-labeling generally fail, as their core assumption of a single, confident ground-truth label is incompatible with the inherent ambiguity of emotional expression. Further, no single method universally excels, and its effectiveness is highly dependent on the distributional shifts and tasks.

\keywords{speech emotion recognition, domain adaptation, test time adaptation, personalization, cross-corpus}
\end{abstract}

%\end{IEEEkeywords}
%

%\vspace{-5pt}
\section{Introduction}
%\vspace{-5pt}
\label{sec:intro}
Emotions are fundamental to human communication, making automatic speech emotion recognition (SER) essential for improving human-computer interaction and health prediction systems~\cite{SER-importance}. Research on SER has shown significant promise in both identifying specific emotional states and tracking dynamic emotional changes over time~\cite{pastor2023cross}. %Existing studies have predominantly concentrated on feature extraction or on designing machine learning models with a primary emphasis on enhancing accuracy~\cite{}.

However, SER performance generally deteriorates when there is a significant domain distribution shift between training and test data, a common challenge in real-world scenarios. For instance, differences in recording devices and data collection procedures can cause substantial variations in speech recordings, leading to distributional changes that negatively impact SER effectiveness~\cite{datacollection-procedure-diffs}. Moreover, SER faces unique distributional challenges compared to other speech-related tasks. Specifically, individuals express emotions differently due to factors such as cultural background and personality, resulting in variability across speakers~\cite{personalized}. This variability causes distribution shifts that hinder SER systems from generalizing to unseen individuals. Furthermore, limitations in data collection often necessitate the use of both acted and natural emotional speech datasets when training machine learning models~\cite{busso2008iemocap}. However, these two types of data exhibit distinct characteristics, introducing additional distributional shifts between them~\cite{act-to-natual-fail}. Therefore, models trained on one type of dataset typically fail to generalize to the other. %Collectively, these issues exacerbate the difficulty of employing SER systems in new and unseen scenarios, 
Addressing these challenges is crucial to making SER systems more reliable, adaptive, and user-friendly when faced with diverse and unpredictable real-world conditions. 

Domain adaptation methods~\cite{DA-1,DA-2} have been widely studied to address such shifts. However, unsupervised domain adaptation typically requires access to source domain data, which is often impractical due to privacy and data-sharing concerns~\cite{privacy-1}. On the other hand, supervised domain adaptation relies on labeled data from the target domain, which is usually unavailable at the test time~\cite{no-label-test-time}. %but they typically require access to the source domain, which is often impractical due to privacy and data-sharing concerns. Fine-tuning~\cite{inductive-TL-1, inductive-TL-2} is another widely explored paradigm, but they requires target data labels which are typically not available at test time. 
To address these challenges, \textit{test-time adaptation} (TTA) offers a compelling alternative, which adapts source model at inference time using only unlabeled target data, without access to source data or target labels. 

Current TTA methods have demonstrated strong potential in image classification and are increasingly being investigated in automatic speech recognition (ASR). These methods generally fall into three categories. \emph{Entropy-minimization approaches} (e.g., TENT~\cite{TENT}, EATA~\cite{EATA}, SAR~\cite{SAR}) adapt models by reducing the prediction uncertainty, encouraging more confident decisions on the target data through backpropagation. \emph{Pseudo-labeling approaches} (e.g., CoTTA~\cite{CoTTA}, AWMC~\cite{AWMC}) generate pseudo-labels, i.e., predicted class labels, for the target data, and use these labels as targets to iteratively refine the model, promoting consistency on the target domain. \emph{Backpropagation-free methods} (BP-free) such as T3A~\cite{T3A}, FOA~\cite{FOA}, LAME~\cite{LAME}, E-BATS~\cite{E-BATS} avoid updating model weights with gradients. Instead, they employ forward-only strategies such as recalibrating the classifier head, %adjusting normalization statistics\jd{do we need to mention this? since we did not experiment TTA methods relate to this}, 
or using prompt-based mechanisms to adapt to new domains. However, the effectiveness of TTA methods in addressing specific distributional shifts in SER such as the discrepancies between acted and natural emotions, remains underexplored.

% Current TTA methods have shown strong promise in image classification and are beginning to be explored in automatic speech recognition (ASR). They can generally be grouped into three categories. Entropy-minimization methods (e.g., TENT~\cite{TENT}, EATA~\cite{EATA}, SAR~\cite{SAR}) adapt the model by increasing prediction confidence through backpropagation in the target domain. Pseudo-labeling methods (e.g., CoTTA~\cite{CoTTA}, AWMC~\cite{AWMC}) update the model based on self-generated\td{predicted?s} labels through backpropagation, aiming to reinforce prediction consistency. Backpropagation-free methods (e.g., T3A~\cite{T3A}, FOA~\cite{FOA}, LAME~\cite{LAME}) avoid costly gradient weight updates and instead adapt the model in a resource-efficient way, for instance by recalibrating the classifier or applying prompt tuning. Given the unique challenges of SER, including speaker variability, acted-versus-natural discrepancies, and corpus-level differences, the effectiveness of existing TTA methods transferred to SER remains unexplored.

%To gain a comprehensive understanding of how TTA addresses various distributional shifts in SER, 
To address this gap, we present the first systematic evaluation and comparison of how TTA methods adapt to a range of distributional shifts in SER. %To capture the diversity of distribution shifts encountered in practice, 
We design three representative and challenging SER tasks: (i) intra-corpus personalization, adapting a universal model to unseen individuals; %within the same corpus; 
(ii) acted-to-natural adaptation, transferring from acted to more naturalistic emotional speech; and (iii) cross-corpus adaptation, transferring between different databases collected under distinct conditions. We evaluate 11 TTA methods spanning entropy-minimization, pseudo-labeling, and backpropagation-free approaches, and evaluated the performance across two commonly used emotion databases. The contributions are summarized as follows: \vspace{-3pt} % and report results at both the category and individual method levelv\. Our findings reveal three key insights: 
\begin{itemize}
    \item  We present the \emph{first comprehensive TTA evaluation for SER}, laying the foundation for the development of improved adaptation techniques across diverse SER tasks. %\vspace{-15pt}
    \item We demonstrated that \emph{backpropagation-free methods are the most promising} among three categories, %showing competitive recognition accuracy across all three tasks, 
    while the most effective specific TTA method still varies across tasks. %\vspace{-5pt}% within the same task. 
    % \item The best-performing individual method in each task also comes from the backpropagation-free category. However, no single method is universally effective across all tasks.
    \item The effectiveness of TTA methods applied with SER is \emph{bounded by type and severity of domain shifts}. % rather than batch reliability.\vspace{-2pt}\td{I just realised we haven't talked about batch before and mentioned it suddently here. maybe just say the former half and remove the latter half?}
    %Entropy-minimization and pseudo-labeling methods are \emph{highly influenced by batch size}, making their performance unstable and less reliable in practice for SER. \jd{Since we did not include batch size analysis, how about we say: Current TTA methods still facing challenges in adapting acted emotions to natrual representations.}\td{a very minor point and may not be the most useful insights.}
\end{itemize}

\section{Methodology}
%\vspace{-5pt}
\subsection{Overview}
%\vspace{-5pt}
Given a source dataset $\mathcal{D}_{s} = \{\bm x_s^{i}, y_s^{i}\}_{i=1}^N$ of $N$ samples, where $\bm x_s^{i}$ denotes the input speech utterance, and $y_s^{i}$ corresponds to an emotion category among $C$ categories. The source model $f_{\bm \theta}$, parameterized by $\bm \theta$, is trained using $D_s$: $\hat{y}_s^i = f_{\bm \theta}(\bm x_s^i)$. Given a target dataset $\mathcal{D}_{t} = \{\bm x_t^{j}\}_{j=1}^M$ of $M$ speech utterances drawn from a different distribution, our objective is to adapt the source model $f_{\bm \theta}$ %to make accurate predictions 
for the target data $y_t^i$ as $\hat{y}_t^j = f_{\bm \theta'}(\bm x_t^j)$.

% We use source and target domain to represent training and test dataset. The test dataset is directly used for adaptation. Let $\mathcal{D_{s}} = \{\bm x_s^{i}, y_s^{i}\}_{i=1}^N$ represent the source dataset, where $\bm x_s^{i}$ denotes the input of speech segment at frame $i$, $y_s^{i}$ corresponds to an emotion category among $C$ total categories, and $N$ is the number of source samples. The source model $f$, parameterized by $\bm \theta$, can be defined using $D_s$: $y_s^i = f(\bm x_s^i \bm \theta)$. Given the target dataset $\mathcal{D_{t}} = \{\bm x_t^{i}\}_{i=1}^N$ containing only speech data with distribution shift, our objective is to refine the source model $f$ to make accurate predictions for the target labels $y_t^i$.

\begin{figure}[t!]
    \centering
    %\vspace{-20pt}
    \includegraphics[width=0.95\linewidth]{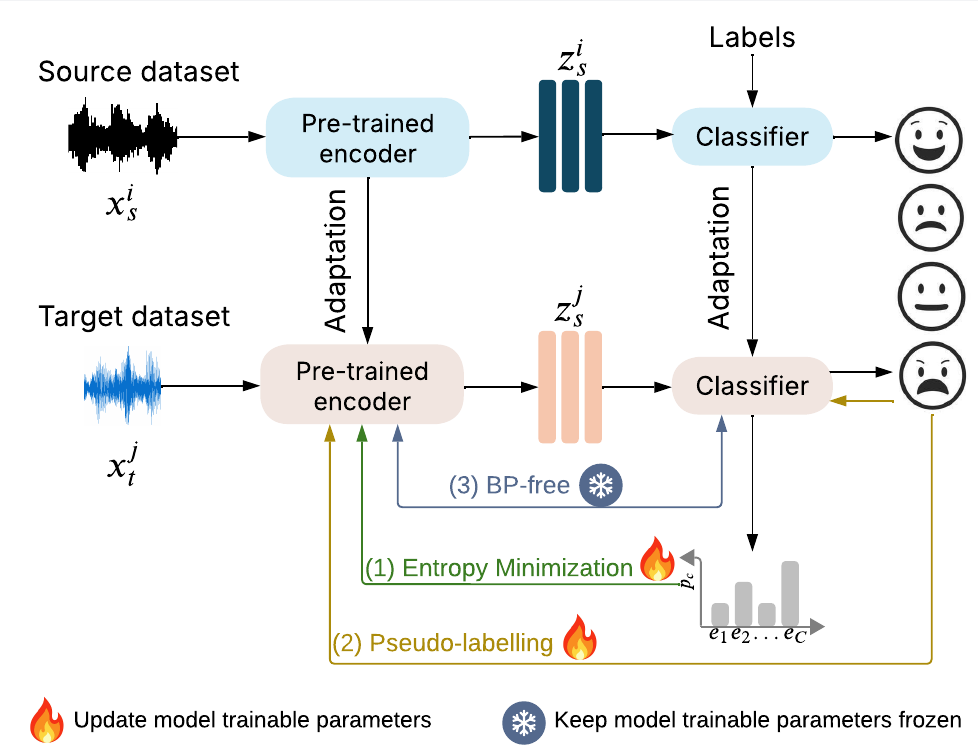}
    %\vspace{-10pt}
    \caption{TTA evaluation for SER. The source model is adapted to the target dataset using only unlabeled speech data.} %\td{label 1,2,3 for each of the method corresponding to the order in 2.2 to 2.4.}}%Entropy minimization, Pseudo-labelling, and BP-free TTA methods are employed for adaptation.}
    %\vspace{-12pt}
    \label{fig:1}
\end{figure}

% \subsection{System overview}
Figure~\ref{fig:1} presents an overview of the TTA evaluation pipeline. During training, a pre-trained encoder processes source speech utterances $\bm{x}_s^i$ and generates latent embeddings $\bm{z}_s^i$, which are then processed by a classifier that assigns each utterance to one of $C$ predefined emotion categories. In the testing phase, target speech utterance $\bm{x}_t^j$ originated from a different distribution from the source data, is encoded. TTA is utilized to update the model parameters %(including either the pre-trained encoder or the classifier) \td{jiaheng, does it include both?} \jd{changed to "or"} 
on-the-fly based solely on the unlabeled target data. In this paper, we employ and compare three types of TTA methods.

% Figure~\ref{fig:1} illustrates the TTA benchmarking pipeline. During training, the pre-trained encoder processes the input speech utterances $\bm x_s^i$ and generates latent embeddings, denoted as $\bm z_s^i$. These embeddings are subsequently passed through a classifier, which categorizes the speech utterances into $C$ predefined emotion classes. In the testing phase, the input speech utterance $\bm x_t^j$ exhibits distributional shift compared to the source speech used during training. Furthermore, only unlabeled speech utterances (i.e., without emotion labels) are available. TTA is employed to adjust the model parameters dynamically based solely on the target data, enabling the system to adapt effectively to the shifted distribution and maintain robust performance under these unseen conditions.
%\vspace{-5pt}
\subsection{Adaptation via Entropy-minimization}
%\vspace{-5pt}
Since test labels are not available, the source model $f_{\bm \theta}$ is adapted to the target dataset in an unsupervised manner. Entropy-minimization~\cite{wangtent} methods update selected trainable parameters (e.g., normalization layers) during inference to increase prediction confidence, by minimizing the prediction entropy for each input as:
%\vspace{-10pt}
\begin{equation}\label{eq:1}
H(p) = - \sum_{c=1}^{C} p^c \log p^c,
%\vspace{-10pt}
\end{equation}
\noindent where $p^c$ denotes the predicted probability for class $c$. %The objective at test time is to 
Minimizing $H(p)$ reduces uncertainty in the model predictions and encourages more confident assignments to a single class, which %This strategy makes the model more decisive under shifted conditions and 
has shown effectiveness in image classification~\cite{TENT,SAR,EATA} and ASR~\cite{SUTA,DSUTA,cea}.

%\vspace{-5pt}
\subsection{Adaptation via Pseudo-labeling }
%\vspace{-5pt}
%Since ground-truth labels are unavailable at test time, pseudo-labeling methods adapt the model by generating labels from its own predictions and enforcing consistency. These methods typically employ two models: a \textit{main model $f_{\theta}$}, which is directly updated, and an \textit{anchor model $f_{\theta'}$}, which updates more smoothly as an exponential moving average (EMA) of the main model. The anchor model provides more reliable pseudo-labels by leveraging both the knowledge from the source model and information accumulated from the target domain. 

Pseudo-labeling methods facilitate model adaptation by leveraging self-generated labels and enforcing prediction consistency~\cite{CoTTA,AWMC}. Typically, these approaches use two models both initialized by the source model $f_{\bm \theta}$: a \emph{main model} $f_{\bm \theta_m}$, which is directly updated during adaptation, and an \emph{anchor model} $f_{\bm \theta_a}$, whose parameters are updated more gradually as the exponential moving average (EMA) of the parameters of the main model $f_{\bm \theta_m}$. The anchor model serves as a stable teacher, producing reliable pseudo-labels that combine source knowledge and accumulated target-domain information. Consistency between the two models predictions helps achieve robust adaptation to distribution shifts.

Given an unlabeled target utterance $\bm{x}_t^j$, the anchor model $f_{\bm \theta_a}$ generates a pseudo label $\hat{y}_{t,a}^j = f_{\bm \theta_a}(\bm{x}_t^j)$, while the main model $f_{\bm \theta_m}$ produces its own prediction $\hat{y}_{t,m}^j = f_{\bm \theta_m}(\bm{x}_t^j)$. To encourage consistency between the two models, the main model is updated by minimizing the cross-entropy loss $\mathcal{L}_{\text{PL}}$ between the anchor model prediction and its prediction:
%\vspace{-10pt}
\begin{equation}\label{eq:pl_loss}
\mathcal{L}_{\text{PL}}(\bm{x}_t^j) = -\sum_{c=1}^C \hat{y}_{t,a}^{j, c} \log \hat{y}_{t,m}^{j, c},
%\vspace{-5pt}
\end{equation}
where $\hat{y}_{t,a}^{j, c}$ and $\hat{y}_{t,m}^{j, c}$ denote the predicted probabilities for class $c$ from the anchor and main models, respectively.

% Given an unlabeled target utterance $\bm x_t^j$, the anchor model $f_{\bm \theta_a}$ produces the pseudo label $\hat{y}_{t,a}^{j} = f_{\bm \theta_a}(\bm x_t^j)$. The main model $f_{\bm \theta_m}$ generates its own prediction $\hat{y}_{t,m}^j = f_{\bm \theta_m}(\bm x_t^j)$. The main model is optimized by minimizing the cross-entropy loss between the pseudo label and its prediction:
% \begin{equation}\label{eq:pl_loss}
% \mathcal{L}_{\text{PL}}(x_t^j) = - \sum_{c=1}^C \hat{y'}_{t,a}^{j} \, \log \hat{y}_{t,m}^j.
% \end{equation}
After the main model updates, the anchor model parameters are updated as an EMA of the main model parameters:
%\vspace{-5pt}
\begin{equation}\label{eq:ema}
\bm \theta_a \leftarrow \gamma \bm \theta_a + (1-\gamma)\bm \theta_m,
%\vspace{-5pt}
\end{equation}
where $\gamma \in [0,1]$ is the momentum controlling the update rate. 

%\vspace{-5pt}
\subsection{Backpropagation-Free Methods}
%\vspace{-5pt}
Backpropagation-free (BP-free) methods adapt the model in a forward-only mannar without gradient computation~\cite{T3A,LAME,FOA,E-BATS}. We highlight three representative BP-free mechanisms: classifer head recalibration, maximum-likelihood estimation (MLE) correction, and forward-only prompt tuning.%\td{the subtitle can be method or mechanism in general, but when you discuss each of them, you can mention the method name in the text. 'Three different mechanisms are proposed for BP-free methods. Prototype xxx. This method aims to xx, which is used in T3A [ref]. ' I skipped this part for now and will read after you revise this part.}

\textit{Classifier head recalibration.} 
This mechanism adjusts the class prototypes directly using the model predictions, as proposed in T3A~\cite{T3A}. The predictions for the unlabeled target samples are first ranked by confidence, and only the most certain ones are selected to serve as reliable pseudo-labels. A new prototype, $\bm{\mu}_c$, is subsequently constructed for each class $c$ by computing the mean of the feature embeddings for these high-confidence target samples assigned that pseudo-label.
Once the prototypes are recalibrated, a target sample latent embedding $\bm{z}_t^{j}$ is classified by assigning it the label of the prototype $\bm{\mu}_c$ that yields the highest softmax probability score as:
% Specifically, it adapts the classifier by constructing class prototype ${\bm{\mu}_c}$ for each class $c$ by averaging the $\bm{z}_t$ of confident target samples of that class. Specifically, it estimate a prototype ${\bm{\mu}_c}$ for each class $c$ by computing the mean vector of the corresponding class-wise embeddings of target samples, where the class labels are taken as pseudo-labels. Confidence ranking is utilized to reduce misleading risk during the classifier update process, only high confidence data samples will be used for updating, ultimately producing more reliable prototypes for improved classification performance. \jd{updated, as for the rule it follows to update the classifier head I think is by only using high confidence samples to update classifier}
% Classification is then performed by taking the argmax over adjusted class prototype:
%\vspace{-10pt}
\begin{equation}
\hat{y}_t^{\,j} = \arg\max_{c \in \{1,\dots,C\}}
\frac{\exp\left(\bm{z}_t^{\,j} \cdot \bm{\mu}_c\right)}
     {\sum_{c'=1}^{C} \exp\left(\bm{z}_t^{\,j} \cdot \bm{\mu}_{c'}\right)} .
\end{equation} 
%\td{vector should be bold, check how it is written in previous sections and revise.} \jd{updated}
%\td{what rule it follows to update the classier head? Prototype is not enough right? It only computes the mean of each cluster, so what is guided to generate more reliable clusters or means? This is somehow missing? Here I assume the key idea is: estimate a prototype for each class by computing the mean vector of its corresponding class-wise embeddings, where the class labels are taken as pseudo-labels. xx samples is utilized to xx during the classifier update process, ultimately producing more reliable prototypes for improved classification performance, or sth similar. }

\textit{Maximum-likelihood estimation (MLE) correction.}  
This also aims to adapt prediction probabilities directly without changing model parameters~\cite{LAME}. %The core principle is to enforce consistency ensuring that inputs with similar latent features receive similar predictions, while regularizing the predictions to remain faithful to the outputs of the source model. 
The core principle is to maximize the likelihood by enforcing the prediction close to the outputs of the source model, and correct it by utilizing Laplacian regularization, which encourages that inputs with similar latent features to receive similar predictions.
The method takes the initial prediction from the source model, $\hat{y}_{t}^j$, and calculates a new, adapted prediction, $\hat{y}'_{t}{}^j$, by finding a value that minimizes a balance objective function:
%\vspace{-10pt}
\begin{equation}
\hat{y}'_{t}{}^j = \arg\min_{y'} \left( \underbrace{\text{dist}(y', y_{\text{neighbors}})}_\text{Consistency} + \lambda \cdot \underbrace{\text{dist}(y', \hat{y}{t}{}^j)}_\text{Source Fidelity} \right)
%\vspace{-5pt}
\end{equation}
The consistency term promotes agreement among similar samples, while the source fidelity term prevents excessive deviation. $\lambda$ balances these two objectives.
% The core idea is to make prediction probability similar for inputs with similar latent features and dissimilar otherwise while stay close to source beliefs. To achieve these using source-domain knowledge, it (i) embeds target samples with the source-trained encoder and constructs an affinity graph (e.g., k-nearest neighbours with distance-based weights), treating nearby points as similar; and (ii) anchors the adjusted probabilities to the source model’s original outputs via a divergence penalty. 
%This mechanism aims to adapt predictions by correcting the output probability of a frozen classifier using neighboring target samples with source domain knowledge, as in LAME~\cite{LAME}. Specifically, it builds an affinity graph $W=[w_{ij}]$ over features $z_t^j$ for a batch of data. Then, LAME minimizes the objective to force data samples with similar latent embedding to provide a similar output probability distribution while staying close to the source model's beliefs: 
%\begin{equation}
%\label{eq:lame-original}
%\arg\min_{\{ p_t^{j}\in\Delta^{C-1}\}}
%\left[
%\sum_{j=1}^{M}\mathrm{KL}\!\big( p_t^{j}\,\|\, q_t^{j}\big)
\;%-\;
%\sum_{i=1}^{M}\sum_{j=1}^{M} w_{ij}\,  p_t^{i\top} p_t^{j}
%\right].
%\end{equation}
%where $KL$ means the Kullback-Leibler divergence~\cite{KL-divergence}, \(q_t^{j}=\mathrm{softmax}\!\big(f_{\bm{\theta}}(\bm{x}_t^{j})\big)\) treats as the model’s original class probabilities, and \( p_t^{j}\) is denoted as the adjusted output probability distribution for \(\bm x_t^{j}\).\td{where is the maximum-likelihood estimation? ' This mechanism aims to adapt prediction probabilities directly, without altering the model parameters [21]. The core idea is to encourage predicted probabilities to be similar for inputs with similar feature representations, and dissimilar otherwise. To evaluate similarity in the feature space as a guideline, source domain knowledge is leveraged by XXX (explain how, i am not sure what this means). This then serves as guidance to align the output probabilities accordingly as: '. Also for equation 5, i, and j here is confusing, as you use i for source domain before, here does i also refer to source domain? But in the text of the affinity graph, you are referring that to a batch of data, and not sure how this aligns. Also in this case, we don't care graph, and we can refer w as the similarity matrix or sth for clarity. If the equation takes too much space to explain, and if it is easier to describe the methods, then you can skip the equations.}

\textit{Forward-only prompt tuning.}  
This aims to introduce learnable prompt embeddings that are directly concatenated with the input to shift activations without requiring backpropagation, as utilized in FOA~\cite{FOA}. These prompts are optimized using a derivative-free method of Covariance Matrix Adaptation Evolution Strategy (CMA-ES), to reduce both prediction uncertainty as in Eq.~\eqref{eq:1} and the discrepancy between source and target domain features. The loss function is defined as: % In addition to entropy loss, it further introduces the discrepancy loss which minimizes the latent embedding distribution differences between target and source domain. The final loss function is:
%\vspace{-5pt}
\begin{equation}
\mathcal{L}_{\text{FOA}} = H({p}) + 
\lVert \bm{\bar{z}}^T - \bm{\bar{z}}^S \rVert_2 
+ \lVert \bm{\sigma}^T - \bm{\sigma}^S \rVert_2 .
%\vspace{-5pt}
\end{equation}
$\bm{\bar{z}}^T$ and $\bm{\bar{z}}^S$ represent the mean feature embeddings for the target and source domains respectively, and $\bm{\sigma}^T$ and $\bm{\sigma}^S$ denote the standard deviations in the target and source domains.

% where $\{\bar{z}^T, \sigma^T\}$ are the target statistics calculated from the latent embeddings of a batch of target data samples. Similarly, the $\{\bar{z}^S, \sigma^S\}$ are calculated based on a precollected small subset of source utterances, which could be used forever once calculated.

%Since test labels are not available, the source model \( f \) is adapted to the target dataset in an unsupervised manner. To facilitate this adaptation without the need for labeled data, entropy minimization~\cite{wangtent} TTA methods update specific layer's trainable parameters to maximize the prediction confidence. The entropy of the model's predictions is defined as:
%\begin{equation}\label{eq:1}
%H(p) = - \sum_{c=1}^{C} p_c \log p_c
%\end{equation}
%\noindent where $p_c$ is the predicted probability for each class $c$. The objective during the inference phase is to minimize \( H(p) \), which reduces the uncertainty in the model's predictions by encouraging it to assign higher confidence to one of the possible classes. By doing so, the model becomes more decisive in its predictions, effectively mitigating the impact of distributional shifts to a certain extent in image classification task and ASR task. 

%\vspace{-5pt}
\section{Experimental setup}
%\vspace{-5pt}
%\paragraph*{Databases.}
\subsection{Databases}
%We use IEMOCAP~\cite{busso2008iemocap} and RAVDESS~\cite{livingstone2018ryerson}. IEMOCAP contains about 12 hours of English dyadic recordings from 10 speakers across 5 sessions, including scripted (acted) and improvised (natural) dialogues. We adopt four emotion classes, happiness with excitation merged, anger, sadness, and neutral, yielding 5{,}531 utterances. For RAVDESS, we use the speech recordings where 24 professional actors produce %two fixed sentences across 
%eight emotions (calm, happy, sad, angry, fear, surprise, disgust, neutral), totaling 1{,}440 utterances. 

We utilized two databases: IEMOCAP~\cite{busso2008iemocap} and RAVDESS~\cite{livingstone2018ryerson}. The IEMOCAP database encompasses 12 hours of audio-visual recordings in English, featuring 10 speakers. These speakers participate in dyadic conversations, with pairs forming 5 distinct sessions. Each session contains scripted (acted) and improvised (natural) dialogues.% between two speakers. 
The acted emotions refer to the emotions portrayed by the actors, while in the improvised part, the actors were asked to recall a past memory and express themselves based on a specific emotion, resulting in more natural emotional expressions. Four emotion categories are employed: happiness, anger, sadness, and neutral. Excitation class is merged with happiness to better balance the size of each emotion class, which results in a total of 5,531 utterances. 

The RAVDESS database contains both speech and song segments, and our study primarily focuses on the speech segments. 24 professional actors (12 females and 12 males) vocalized two predetermined sentences across different emotion states. The database annotates eight emotion states: calmness, happiness, sadness, anger, fear, surprise, disgust, and a neutral expression, resulting in a total of eight emotions and a total of 1440 utterances. %\td{if no space, this part can be cut.}
%\vspace{-5pt}
%\paragraph*{Source Model Structure and Training Settings. }
\subsection{Source Model Structure and Training Settings}
Speech is first segmented into 5-second and 8-second windows for IEMOCAP and RAVDESS respectively. Wav2Vec 2.0 ~\cite{baevski2020wav2vec}\footnote{\href{https://huggingface.co/facebook/wav2vec2-base}{https://huggingface.co/facebook/wav2vec2-base}} is used as the encoder to extract the features. Linear projection with a single fully connected layers of 256 neuros was employed for classification. The model is initially fine-tuned for SER in the source domain using the AdamW optimizer for 50 epochs with an optimized learning rate of 3e-5. A linear warm-up is applied over 10\% of the training steps, followed by a linear decay scheduler. Performance is evaluated using accuracy and macro F1 score.

%\td{The model is trained for xxx epochs with the xxx optimiser. The learning rate is initialized to xxx, with a decay factor of xxx. Performance evaluation is based on the macro-averaged F1 score to accommodate for class imbalances.}

%\td{Hyperparameters iemocap: lr 8.28e-05, margin: 0.10358546771410294 RavDess: lr: 4.4670457365805335e-05, margin: 0.23600459270848917}
%\td{baselines}in latex
%\vspace{-5pt}
%\paragraph*{Implementation Details of TTA. }
\subsection{Implementation Details of TTA}
%All TTA methods operate on unlabeled target batches and adapt the source model at inference time. %Unless otherwise noted, 
We follow the original settings of each TTA method\footnote{Code is available at \href{https://github.com/JiahengDong/SETTA}{https://github.com/JiahengDong/SETTA}}. %for which parameters are updated (e.g., normalization layers, feature extraction layers) and the adaptation steps for each utterance. 
For fairness, hyperparameters were chosen by grid search. As for Entropy minimization and Pseudo-labelling methods, we use AdamW with learning rate of $1e^{-5}$ for backpropagation. The reported results for each task is based on batch size of 32. We analyses the TTA performance across different batch sizes within section~\ref{sec:batch-size-analysis}.%\td{As you already have the batch size results, here change it to: We analyses the TTA performance across different batch sizes within [xxx]}. Then in the performance section, you can briefly mention that the results is based on batch size of 32. } %As for TTA methods (SUTA, DSUTA, AWMC, CEA) which are designed for single utterance adaptation, we use batch size of 1, others we evaluate with batch size of 32. \td{so the comparison of baselines in the results part use different batch sizes? Not fair as you somehow changed the task settings already.}%To further test the impact of the batch size for TTA in SER, we investigated different batch sizes {1, 16, 32, 64}, and the adaptation results are reported with average accuracy score\td{across all batches? why} and standard deviation.

%\vspace{-5pt}
%\paragraph*{Tasks. } 
\subsection{Tasks}
We evaluate the model on three SER tasks:
%\begin{enumerate}
%    \item \emph{Intra-corpus personalization}, where the model is adapted to new individual speaker. In IEMOCAP, we use leave-one-session-out, and adaptation was carried out for each individual. For RAVDESS, 20 subjects are used for training and the remaining four are used for testing.
%    \item \emph{Acted-to-Natural Adaptation}, which adapts a model trained on scripted, acted emotions to recognize more natural, improvised ones, using IEMOCAP given the data availability.
%    \item \emph{Cross-corpus generalization}, where model trained on one dataset is adapted to the other, using the four shared emotion categories across datasets.
%\end{enumerate}
%\emph{i) Intra-corpus personalization}, where the model is adapted to new individual speaker. In IEMOCAP, we use leave-one-session-out, and adaptation was carried out for each individual. For RAVDESS, 20 subjects are used for training and the remaining four are used for testing. \emph{ii) Acted-to-Natural Adaptation}, which adapts a model trained on scripted, acted emotions to recognize more natural, improvised ones, using IEMOCAP given the data availability. \emph{iii) Cross-corpus generalization}, where model trained on one dataset is adapted to the other, using the four shared emotion categories across datasets.
\begin{itemize}[leftmargin=0.11in]
     \item \textbf{Task 1: Intra-corpus personalization}: This task focused on adapting the source model to individual users, aiming to develop a personalized model. In IEMOCAP, the experiments were carried out using leave-one-session-out, and adaptation within the corpus was carried out at individual levels. i.e., adapting to each individual in the held-out session, and in total two adaptation was carried out for each fold. For RAVDESS dataset, 20 subjects are used for training and the remaining four are used for testing.  
     \item \textbf{Task 2: Acted to natural emotion adaptation}: This task adapts a model trained on scripted dialogues with acted emotions to improvised dialogues of natural emotional expressions using IEMOCAP given its data availability.
     \item \textbf{Task 3: Cross-corpus generalization}: This task involves adapting the source model, originally trained on one database, to another dataset. For example, adapting a model trained on IEMOCAP to RAVDESS, and vice versa. The adaptation adopts four emotional categories shared between two datasets.
\end{itemize}

%\vspace{-5pt}
\section{Results}
%\vspace{-5pt}
\begin{table*}[t]
\centering
\caption{Recognition accuracy (\%) and F1 (\%) across three tasks.
\textbf{Bold} = best TTA method; \underline{underline} = best average. %Methods: Tent~\cite{TENT}, SAR~\cite{SAR}, EATA~\cite{EATA}, SUTA~\cite{SUTA}, DSUTA~\cite{DSUTA}, CEA~\cite{cea}, CoTTA~\cite{CoTTA}, AWMC~\cite{AWMC}, T3A~\cite{T3A}, LAME~\cite{LAME}, FOA~\cite{FOA}.
}

\label{tab:tta-3-in-row-big}
\setlength{\tabcolsep}{4pt}
\renewcommand{\arraystretch}{1.05}
\newlength{\panelgap}
\setlength{\panelgap}{1.0 em} % <-- adjust to taste
% -------- (a) TASK 1 --------
\begin{minipage}[t]{0.28\textwidth}
\centering\footnotesize
\textbf{(a) Task 1: IEMOCAP \& RAVDESS}\par\vspace{2pt} \label{tab:task1_avg_results}
\begin{adjustbox}{max width=\linewidth}
\begin{tabular}{lcc|cc}
\toprule
\textbf{Method} & \multicolumn{2}{c}{IEMOCAP} & \multicolumn{2}{c}{RAVDESS} \\
 & Acc & F1 & Acc & F1 \\
\midrule
Source model & 67.1 & 67.4 & 72.5 & 69.5 \\
\midrule
\multicolumn{5}{l}{\textit{Entropy-minimization (EM)}} \\
Tent~\cite{TENT}  & 66.6 & 66.9 & 72.5 & 69.5 \\
SAR~\cite{SAR}   & 67.1 & 67.4 & 72.5 & 69.5 \\
EATA~\cite{EATA}  & 66.6 & 66.8 & 72.5 & 69.5 \\
SUTA~\cite{SUTA}  & 66.6 & 66.7 & 70.7 & 68.0 \\
DSUTA~\cite{DSUTA} & 66.6 & 66.9 & 70.8 & 68.0 \\
CEA~\cite{cea}   & 66.6 & 66.8 & 73.3 & 70.3 \\ \cmidrule{2-5}
\textit{EM Avg}& 66.7 & 66.9 & 72.1 & 69.1 \\
\midrule
\multicolumn{5}{l}{\textit{Pseudo-labeling (PL)}} \\
CoTTA~\cite{CoTTA} & 66.7 & 67.0 & 69.6 & 66.4 \\
AWMC~\cite{AWMC}  & 67.1 & 67.4 & 72.5 & 69.5 \\ \cmidrule{2-5}
\textit{PL Avg}& 66.9 & 67.2 & 71.1 & 68.0 \\
\midrule
\multicolumn{5}{l}{\textit{BP-free}} \\
T3A~\cite{T3A}   & 67.1 & 67.4 & 72.5 & 69.5 \\
LAME~\cite{LAME}  & 67.2 & 67.7 & 70.4 & 67.1 \\
FOA~\cite{FOA}   & \textbf{67.6} & \textbf{68.2} & \textbf{73.8} & \textbf{70.9} \\ \cmidrule{2-5}
\textit{BP-free Avg} & \underline{67.3} & \underline{67.8} & \underline{72.2} & \underline{69.2} \\
\bottomrule
\end{tabular}
\end{adjustbox}
\end{minipage}
\hspace{\panelgap} %\vspace{-10pt}
% -------- (b) TASK 2 --------
\begin{minipage}[t]{0.193\textwidth}
\centering\footnotesize
\textbf{(b) Task 2: IEMOCAP}\par\vspace{2pt}\label{tab:task2_avg_results}
\begin{adjustbox}{max width=\linewidth} % avoid upscaling the font
\begin{tabular}{lcc}
\toprule
\textbf{Method} & \multicolumn{2}{c}{IEMOCAP} \\
 & Acc & F1 \\
\midrule
Source model & 51.3 & 51.0 \\
\midrule
\multicolumn{3}{l}{\textit{Entropy-minimization (EM)}} \\
Tent  & 50.8 & 50.6 \\
SAR   & \textbf{51.5} & 51.0 \\
EATA  & 51.3 & 51.0 \\
SUTA  & 48.3 & 47.9 \\
DSUTA & 46.1 & 46.4 \\
CEA   & 49.9 & 49.7 \\ \cmidrule{2-3}
\textit{EM Avg} & 49.7 & 49.4 \\
\midrule
\multicolumn{3}{l}{\textit{Pseudo-labeling (PL)}} \\
CoTTA & 51.1 & 50.8 \\
AWMC  & 51.3 & 51.0 \\ \cmidrule{2-3}
\textit{PL Avg} & 51.2 & 50.9 \\
\midrule
\multicolumn{3}{l}{\textit{BP-free}} \\
T3A   & 51.1 & 50.9 \\
LAME  & \textbf{51.5} & \textbf{51.3} \\
FOA   & 51.3 & 51.1 \\ \cmidrule{2-3}
\textit{BP-free Avg} & \underline{51.3} & \underline{51.0} \\
\bottomrule
\end{tabular}
\end{adjustbox}
\end{minipage}
\hspace{\panelgap}
% -------- (c) TASK 3 --------
\begin{minipage}[t]{0.28\textwidth}
\centering\footnotesize
\textbf{(c) Task 3: RAVDESS \& IEMOCAP}\par\vspace{2pt} \label{tab:task3_avg_results}
\begin{adjustbox}{max width=\linewidth}
\begin{tabular}{lcc|cc}
\toprule
\textbf{Method} & \multicolumn{2}{c}{RAVDESS} & \multicolumn{2}{c}{IEMOCAP} \\
 & Acc & F1 & Acc & F1 \\
\midrule
Source model & 37.8 & 26.7 & 50.0 & 45.7 \\
\midrule
\multicolumn{5}{l}{\textit{Entropy-minimization (EM)}} \\
Tent  & 37.4 & 26.2 & 50.0 & 45.6 \\
SAR   & 37.8 & 26.7 & 50.0 & 45.7 \\
EATA  & 37.8 & 26.7 & 49.9 & 45.7 \\
SUTA  & 31.1 & 16.3 & 48.9 & 45.5 \\
DSUTA & 30.5 & 14.9 & 50.1 & 45.8 \\
CEA   & 31.6 & 17.2 & 50.1 & 46.1 \\ \cmidrule{2-5}
\textit{EM Avg}& 34.4 & 21.3 & 49.8 & \underline{45.7} \\
\midrule
\multicolumn{5}{l}{\textit{Pseudo-labeling (PL)}} \\
CoTTA & 32.7 & 19.0 & 50.0 & 45.6 \\
AWMC  & 37.8 & 26.7 & 50.0 & 45.7 \\ \cmidrule{2-5}
\textit{PL Avg}& 35.3 & 22.9 & \underline{50.0} & \underline{45.7} \\
\midrule
\multicolumn{5}{l}{\textit{BP-free}} \\
T3A   & \textbf{43.8} & \textbf{34.3} & \textbf{50.2} & \textbf{46.2} \\
LAME  & 28.7 & 12.3 & 49.6 & 45.2 \\
FOA   & 40.9 & 30.3 & 49.4 & 45.3 \\ \cmidrule{2-5}
\textit{BP-free Avg} & \underline{37.8} & \underline{25.6} & 49.7 & 45.2 \\
\bottomrule
\end{tabular}
\end{adjustbox}
\end{minipage}
\end{table*}

\subsection{Intra-corpus Personalization}
%\vspace{-5pt}
%Table \ref{tab:task1_avg_results} shows the average accuracy of intra-corpus personalization on IEMOCAP and RAVDESS. BP-free methods achieved the highest average accuracy on IEMOCAP (67.3\%) and RAVDESS (72.2\%). 
Table~\ref{tab:task1_avg_results}(a) reports the accuracy and F1 score of intra-corpus personalization. % on IEMOCAP and RAVDESS. 
BP-free methods achieved the highest average performance on both datasets. %with 67.3\% and 67.8\% in terms of accuracy and F1 score on IEMOCAP, and 72.2\% and  69.2\% on RAVDESS. 
The superior performance over entropy minimization and pseudo-labeling TTA may stem from their reduced susceptibility to catastrophic forgetting. Freezing the backbone helps preserve source knowledge, which is advantageous with subtle distribution shifts, as in intra-corpus personalization. Moreover, entropy minimization and pseudo-labeling methods assume a single true class and maximize its probability, which is less suited to SER, where emotions are ambiguous and one-label ground truths may not reflect underlying emotion~\cite{emotion-multilabel}. %This reliance on a single label can be misleading during adaptation and may further degrade performance. 
While BP-free methods may inevitably inherit this assumption to some extent, they also incorporate additional mechanisms, such as discrepancy loss % and \jd{MLE adjustment}, 
that align distributions in a more flexible manner without relying as heavily on this assumption.
%BP-free methods are potentially better since they typically keep the backbone frozen avoids catastrophic forgetting and preserves reliable source knowledge, which remains particularly valuable when adapting to individual users with subtle differences.

%Among different BP-free methods, \textbf{FOA} performed best overall, reaching 67.6\% on IEMOCAP (+0.5\% over the baseline) and 73.8\% on RAVDESS (+1.3\%). 
Among BP-free approaches, FOA performs best. %reaching 67.6\% and 68.2\% F1 on IEMOCAP (+0.5\% / +0.8\% over the baseline) and 73.8\% accuracy / 70.9\% F1 on RAVDESS (+1.3\% / +1.4\%). 
The improvements from FOA likely result from i) its prompt tuning with ii) a discrepancy-based objective. %, which distinguishes it from other BP-free methods. 
Aligning latent features between source and target domains helps maintain a consistent feature space, reducing distribution shifts more reliably with source references. Such an approach might be particularly beneficial for personalization, where subtle speaker variability manifests more clearly in latent features~\cite{speaker-specific-latent,speaker-specific-featuer-eachlayer}. %In contrast, T3A and LAME focus on modifying the classifier layer based on predictions, which may be less effective when the domain shift is subtle in the prediction probability distribution.

%\vspace{-5pt}
\subsection{Performance on Acted-to-natural Adaptation}
%task2 -- fine-tuned on IEMOCAP-scripted -- tested on improved -- target size 2943
%\vspace{-5pt}
As shown in Table~\ref{tab:task2_avg_results}(b), BP-free methods still obtained the best performance, with LAME performed best. %slightly higher than pseudo-labeling %(51.2\% / 50.9\%) 
%and clearly better than entropy-minimization. %(45.1\% / 44.5\%). LAME achieved the best performance. %(51.5\% accuracy / 51.3\% F1), 
%outperforming the source model baseline %(51.3\% / 51.0\%) 
%and other BP-free approaches. 
However, the improvement is minimal, as with most other TTA methods.

These findings highlight the limited effectiveness of current TTA techniques for acted-to-natural adaptation. % under the more challenging shift from acted to natural emotional speech. 
We hypothesize that the domain shift between acted and natural emotions is too complex and inconsistent for current TTA methods to model effectively. TTA typically excels when domain shifts are uniform and global, such as the consistent blurs or brightness changes applied to image datasets~\cite{Imagenet-C}. In contrast, the shift from acted to natural emotions is far more nuanced. For instance, in IEMOCAP, acted emotions are often characterized by subtler acoustic cues tied to fixed transcripts, whereas improvised emotions are typically more intense and expressed more freely~\cite{busso2008iemocap, diffs_act_natrual}. This variability creates a complex, non-uniform shift pattern that current TTA methods are not equipped to capture. 
\subsection{Performance on Cross-corpus Adaptation}
%\vspace{-5pt}
As shown in Table~\ref{tab:task3_avg_results}(c), T3A reached the best performance on both datasets, showing a 3.1\% and 4.1\% average gain for accuracy and F1 score over the source baselines. This is likely because when the source model performance is weak, it generates a high proportion of low-confidence predictions. By recalibrating the prototype using only the most confident samples, the method effectively filters out this noise, preventing the numerous uncertain predictions from biasing the new prototype and thus creating a more robust classifier.
Overall, the improvements in task 3 are significantly greater than those in tasks 1 and 2 when using the best methods, suggesting that TTA might be more effective at adapting medium to severe distribution shifts compared to minor ones in SER. %On RAVDESS, BP-free methods also achieved the best average performance. % (37.8\% and 25.6\% for accuracy and F1 score), outperforming both entropy-minimization %(34.1\% / 20.8\%) 
\subsection{Analysis of Batch Size} \label{sec:batch-size-analysis} 
\vspace{-4pt}
\begin{figure}[h]
\centering
\begin{subfigure}{0.48\linewidth}
  \centering
  %\vspace{-20pt}
  \includegraphics[width=\linewidth]{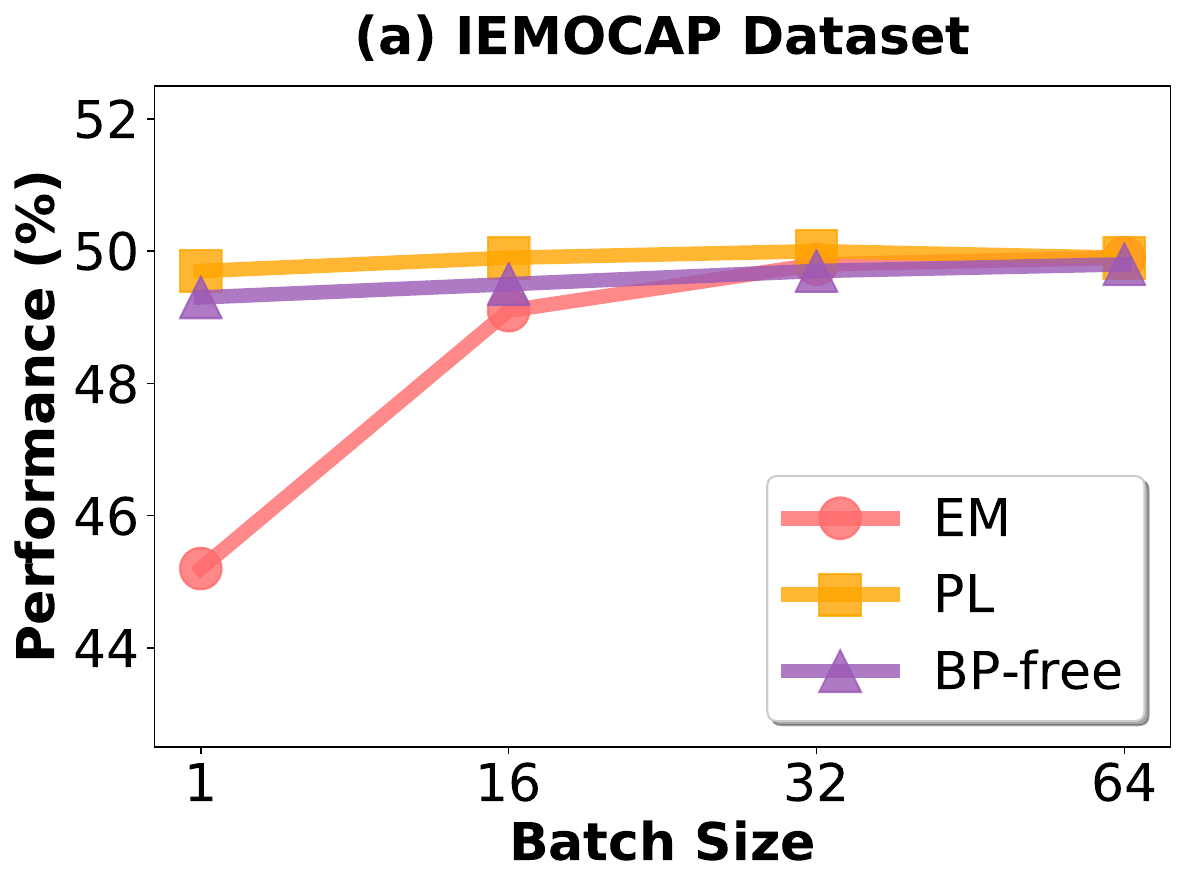}
  \label{fig:batch_task2}
\end{subfigure}\hfill
\begin{subfigure}{0.48\linewidth}
  \centering
  \includegraphics[width=\linewidth]{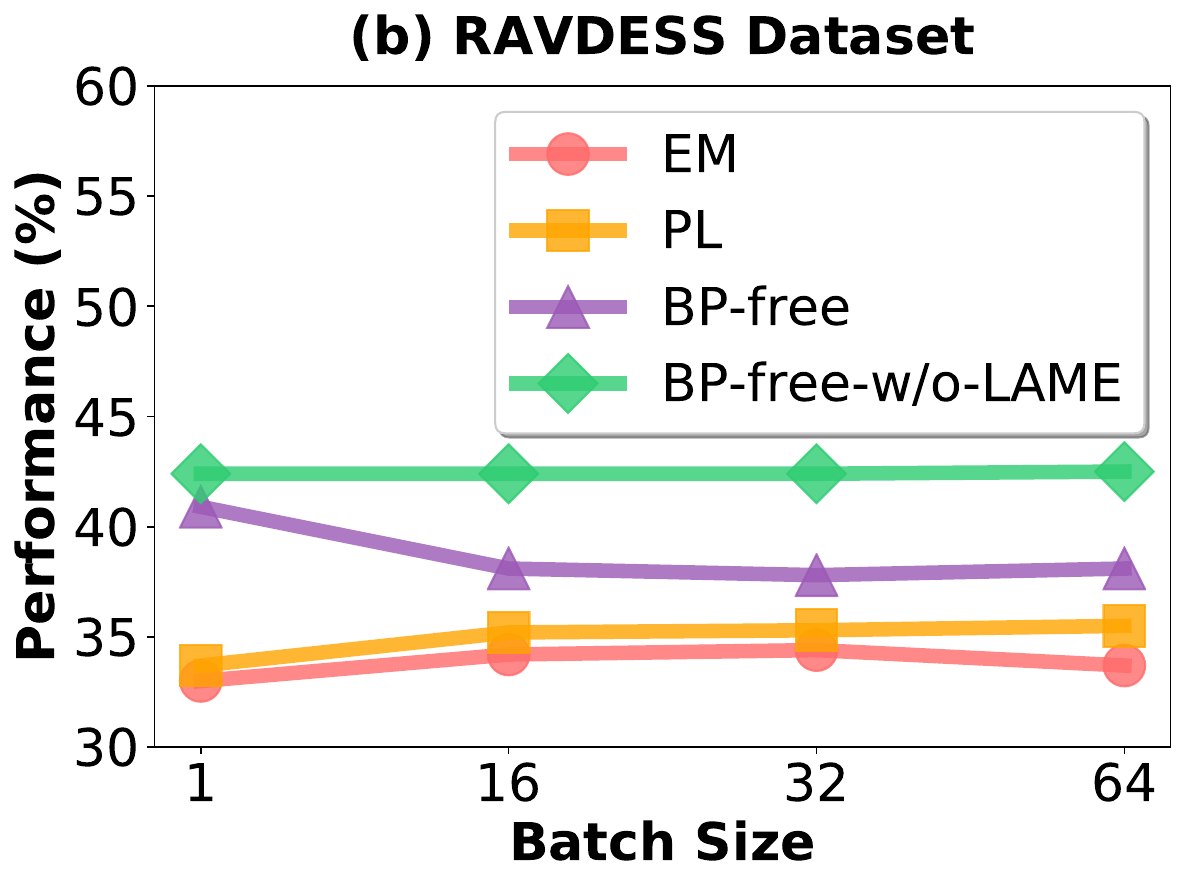}
  \label{fig:batch_task3}
\end{subfigure}
\vspace{-10pt}
\caption{Category average accuracy with different batch size for Task 3 (a) IEMOCAP and (b) RAVDESS.}
\vspace{-10pt}
\label{fig:batch_analysis}
\end{figure}
% \jd{updated both in RAVDESS figure and the analysis. Morevoer, i think the table results is align with these figures, since in Task 3 the average performance on IEMOCAP, they are very close, but PL is actually better, as in the results analysis, we just said T3A performs all the best and have large gains. As for BP-free methods, we just mentioned that they achieved the best on RAVDESS.}
We further analyzed the impact of batch size in TTA on the IEMOCAP and RAVDESS datasets, as shown in Figure~\ref{fig:batch_analysis}. In IEMOCAP, PL and BP-free methods are less affected by batch size, whereas EM is significantly impacted, where a small batch size of 1 can result in unreliable statistics for updates, as demonstrated in~\cite{SAR}. On the RAVDESS dataset, the performance of PL and EM methods improves with increasing batch size, which typically leads to more reliable and stable updates. Interestingly, BP-free performance decreases as batch size increases. This is likely due to the significantly poorer performance when using LAME, which builds a neighbourhood graph from the target batch to enforce consistency. With severe domain shifts, larger batches introduce more misclassifications and unstable corrections. The average performance of the BP-free method without LAME is superior and less affected by batch size. %which uses all the data from a batch, the assumption of data samples with closer distance of feature embeddings should have similar output probability might not hold in this dataset.
With batch size from 16 to 64, the general performance remains stable for all TTA methods. % which also indicates that the bottleneck of the effectiveness of applying current TTA methods in SER is more about the type and severity of domain shifts.%\td{Here in IEMOCAP, PL seems performing best, not consistent with table 3? } increasing batch size from 1 to 64 makes the curves smoother (less variance) but shows marginal or no improvements from batch size 16 to 64. This plateau suggests that the bottleneck in applying current TTA methods in SER is not about the reliability of the batch size but the task and shifts of SER themselves. %In other words, simply scaling batch size does not unlock further adaptation; effectiveness is bounded by shift characteristics and complexity rather than batch reliability. 

\section{Conclusion}
This paper provides the first systematic evaluation and comparison of test-time adaptation (TTA) for SER, evaluating a suite of state-of-the-art methods across three representative SER tasks. Our findings reveal that while BP-free TTA methods show the most promise, no single approach is universally superior. Instead, the optimal method depends on the nature of the distributional shift, with methods like FOA, LAME, and T3A each excelling in different scenarios. We observe that TTA yields significant improvements in cross-corpus settings but provides only marginal gains for cases such as adapting from acted to natural emotions. Moreover, increasing batch size could produce more stable performance, but the improvement is marginal with batch size increase, excluding the extremely small batch size. This study provides a foundational investigation and offers critical insights for future research into more robust and adaptive SER systems.
 
 %Moreover, the bottleneck of TTA methods performance is not the batch size reliability but the type and severity of domain shifts in SER. \jd{updated the final sentence.}%\Tilde{how do we demonstrate this? have we quantified any domain shift characteristics? This is not convincing without support.}

% \begin{figure}[htb]

% \begin{minipage}[b]{1.0\linewidth}
%   \centering
%   \centerline{\includegraphics[width=8.5cm]{Images/image1.eps}}
% %  \vspace{2.0cm}
%   \centerline{(a) Result 1}\medskip
% \end{minipage}
% %
% \begin{minipage}[b]{.48\linewidth}
%   \centering
%   \centerline{\includegraphics[width=4.0cm]{Images/image3.eps}}
% %  \vspace{1.5cm}
%   \centerline{(b) Results 3}\medskip
% \end{minipage}
% \hfill
% \begin{minipage}[b]{0.48\linewidth}
%   \centering
%   \centerline{\includegraphics[width=4.0cm]{Images/image4.eps}}
% %  \vspace{1.5cm}
%   \centerline{(c) Result 4}\medskip
% \end{minipage}
% %
% \caption{Example of placing a figure with experimental results.}
% \label{fig:res}
% %
% \end{figure}

\newpage
\begin{small}
\bibliographystyle{IEEEtran}
\bibliography{strings,refs}

@article{baevski2020wav2vec,
  title={wav2vec 2.0: A framework for self-supervised learning of speech representations},
  author={Baevski, Alexei and Zhou, Yuhao and Mohamed, Abdelrahman and Auli, Michael},
  journal={Advances in neural information processing systems},
  volume={33},
  pages={12449--12460},
  year={2020}
}

@article{busso2008iemocap,
  title={IEMOCAP: Interactive emotional dyadic motion capture database},
  author={Busso, Carlos and Bulut, Murtaza and Lee, Chi-Chun and Kazemzadeh, Abe and Mower, Emily and Kim, Samuel and Chang, Jeannette N and Lee, Sungbok and Narayanan, Shrikanth S},
  journal={Language resources and evaluation},
  volume={42},
  pages={335--359},
  year={2008},
  publisher={Springer}
}

@article{livingstone2018ryerson,
  title={The Ryerson Audio-Visual Database of Emotional Speech and Song (RAVDESS): A dynamic, multimodal set of facial and vocal expressions in North American English},
  author={Livingstone, Steven R and Russo, Frank A},
  journal={PloS one},
  volume={13},
  number={5},
  pages={e0196391},
  year={2018},
  publisher={Public Library of Science San Francisco, CA USA}
}

@article{TENT,
  title={Tent: Fully test-time adaptation by entropy minimization},
  author={Wang, Dequan and Shelhamer, Evan and Liu, Shaoteng and Olshausen, Bruno and Darrell, Trevor},
  journal={arXiv preprint arXiv:2006.10726},
  year={2020}
}

@inproceedings{EATA,
  title={Efficient test-time model adaptation without forgetting},
  author={Niu, Shuaicheng and Wu, Jiaxiang and Zhang, Yifan and Chen, Yaofo and Zheng, Shijian and Zhao, Peilin and Tan, Mingkui},
  booktitle={International conference on machine learning},
  pages={16888--16905},
  year={2022},
  organization={PMLR}
}

@article{SAR,
  title={Towards stable test-time adaptation in dynamic wild world},
  author={Niu, Shuaicheng and Wu, Jiaxiang and Zhang, Yifan and Wen, Zhiquan and Chen, Yaofo and Zhao, Peilin and Tan, Mingkui},
  journal={arXiv preprint arXiv:2302.12400},
  year={2023}
}

@article{FOA,
  title={Test-time model adaptation with only forward passes},
  author={Niu, Shuaicheng and Miao, Chunyan and Chen, Guohao and Wu, Pengcheng and Zhao, Peilin},
  journal={arXiv preprint arXiv:2404.01650},
  year={2024}
}

@article{SUTA,
  title={Listen, adapt, better wer: Source-free single-utterance test-time adaptation for automatic speech recognition},
  author={Lin, Guan-Ting and Li, Shang-Wen and Lee, Hung-yi},
  journal={arXiv preprint arXiv:2203.14222},
  year={2022}
}

@inproceedings{CoTTA,
  author    = {Q. Wang and O. Fink and L. Van Gool and et al.},
  title     = {Continual Test-Time Domain Adaptation},
  booktitle = {Proceedings of the IEEE/CVF Conference on Computer Vision and Pattern Recognition},
  year      = {2022},
  pages     = {7201--7211}
}

@article{DSUTA,
  title={Continual Test-time Adaptation for End-to-end Speech Recognition on Noisy Speech},
  author={Lin, Guan-Ting and Huang, Wei-Ping and Lee, Hung-yi},
  journal={arXiv preprint arXiv:2406.11064},
  year={2024}
}

@inproceedings{AWMC,
  title={Awmc: Online test-time adaptation without mode collapse for continual adaptation},
  author={Lee, Jae-Hong and Kim, Do-Hee and Chang, Joon-Hyuk},
  booktitle={2023 IEEE Automatic Speech Recognition and Understanding Workshop (ASRU)},
  pages={1--8},
  year={2023},
  organization={IEEE}
}

@article{cea,
  title={Advancing Test-Time Adaptation in Wild Acoustic Test Settings},
  author={Liu, Hongfu and Huang, Hengguan and Wang, Ye},
  journal={arXiv preprint arXiv:2310.09505},
  year={2023}
}

@article{T3A,
  title={Test-time classifier adjustment module for model-agnostic domain generalization},
  author={Iwasawa, Yusuke and Matsuo, Yutaka},
  journal={Advances in Neural Information Processing Systems},
  volume={34},
  pages={2427--2440},
  year={2021}
}

@inproceedings{LAME,
  title={Parameter-free online test-time adaptation},
  author={Boudiaf, Malik and Mueller, Romain and Ben Ayed, Ismail and Bertinetto, Luca},
  booktitle={Proceedings of the IEEE/CVF Conference on Computer Vision and Pattern Recognition},
  pages={8344--8353},
  year={2022}
}

@article{DA-1,
  title={Subspace identification for multi-source domain adaptation},
  author={Li, Zijian and Cai, Ruichu and Chen, Guangyi and Sun, Boyang and Hao, Zhifeng and Zhang, Kun},
  journal={Advances in Neural Information Processing Systems},
  volume={36},
  pages={34504--34518},
  year={2023}
}

@article{DA-2,
  title={Hyperdomainnet: Universal domain adaptation for generative adversarial networks},
  author={Alanov, Aibek and Titov, Vadim and Vetrov, Dmitry P},
  journal={Advances in Neural Information Processing Systems},
  volume={35},
  pages={29414--29426},
  year={2022}
}

@INPROCEEDINGS{act-to-natual-fail,
  author={Lashkarashvili, Nineli and Wu, Wen and Sun, Guangzhi and Woodland, Philip C.},
  booktitle={ICASSP 2024 - 2024 IEEE International Conference on Acoustics, Speech and Signal Processing (ICASSP)}, 
  title={Parameter Efficient Finetuning for Speech Emotion Recognition and Domain Adaptation}, 
  year={2024},
  volume={},
  number={},
  pages={10986-10990},
  doi={10.1109/ICASSP48485.2024.10446272}}

@article{privacy-1,
  title={Stress measurement using speech: Recent advancements, validation issues, and ethical and privacy considerations},
  author={Slavich, George M and Taylor, Sara and Picard, Rosalind W},
  journal={Stress},
  volume={22},
  number={4},
  pages={408--413},
  year={2019},
  publisher={Taylor \& Francis}
}

@article{personalized,
  title={Personalized adaptation with pre-trained speech encoders for continuous emotion recognition},
  author={Tran, Minh and Yin, Yufeng and Soleymani, Mohammad},
  journal={arXiv preprint arXiv:2309.02418},
  year={2023}
}

@inproceedings{no-label-test-time,
  title={Tipi: Test time adaptation with transformation invariance},
  author={Nguyen, A Tuan and Nguyen-Tang, Thanh and Lim, Ser-Nam and Torr, Philip HS},
  booktitle={Proceedings of the IEEE/CVF Conference on Computer Vision and Pattern Recognition},
  pages={24162--24171},
  year={2023}
}

@inproceedings{emotion-multilabel,
  title={Speech emotion recognition based on multi-label emotion existence model.},
  author={Ando, Atsushi and Masumura, Ryo and Kamiyama, Hosana and Kobashikawa, Satoshi and Aono, Yushi},
  booktitle={INTERSPEECH},
  pages={2818--2822},
  year={2019}
}

@inproceedings{speaker-specific-latent,
  title={Improving speaker-independent speech emotion recognition using dynamic joint distribution adaptation},
  author={Lu, Cheng and Zong, Yuan and Lian, Hailun and Zhao, Yan and Schuller, Bj{\"o}rn W and Zheng, Wenming},
  booktitle={ICASSP 2024-2024 IEEE International Conference on Acoustics, Speech and Signal Processing (ICASSP)},
  pages={11696--11700},
  year={2024},
  organization={IEEE}
}

@article{speaker-specific-featuer-eachlayer,
  title={Probing Speaker-specific Features in Speaker Representations},
  author={Chiu, Aemon Yat Fei and Fung, Paco Kei Ching and Li, Roger Tsz Yeung and Li, Jingyu and Lee, Tan},
  journal={arXiv preprint arXiv:2501.05310},
  year={2025}
}

@article{Imagenet-C,
  title={Benchmarking neural network robustness to common corruptions and perturbations},
  author={Hendrycks, Dan and Dietterich, Thomas},
  journal={arXiv preprint arXiv:1903.12261},
  year={2019}
}

@article{SER-importance,
  title={Emotion recognition for human--computer interaction using high-level descriptors},
  author={Singla, Chaitanya and Singh, Sukhdev and Sharma, Preeti and Mittal, Nitin and Gared, Fikreselam},
  journal={Scientific reports},
  volume={14},
  number={1},
  pages={12122},
  year={2024},
  publisher={Nature Publishing Group UK London}
}

@article{datacollection-procedure-diffs,
  title={Improving cross-corpus speech emotion recognition with adversarial discriminative domain generalization (ADDoG)},
  author={Gideon, John and McInnis, Melvin G and Provost, Emily Mower},
  journal={IEEE Transactions on Affective Computing},
  volume={12},
  number={4},
  pages={1055--1068},
  year={2019},
  publisher={IEEE}
}

@inproceedings{diffs_act_natrual,
  title={Scripted dialogs versus improvisation: lessons learned about emotional elicitation techniques from the IEMOCAP database.},
  author={Busso, Carlos and Narayanan, Shrikanth S},
  booktitle={Interspeech},
  pages={1670--1673},
  year={2008}
}

@article{E-BATS,
  title={E-BATS: Efficient Backpropagation-Free Test-Time Adaptation for Speech Foundation Models},
  author={Dong, Jiaheng and Jia, Hong and Chatterjee, Soumyajit and Ghosh, Abhirup and Bailey, James and Dang, Ting},
  journal={arXiv preprint arXiv:2506.07078},
  year={2025}
}

@article{wangtent,
  title={tent: fully test-time adaptation by entropy minimization},
  author={Wang, Dequan and Shelhamer, Evan and Liu, Shaoteng and Olshausen, Bruno and Darrell, Trevor}
}

@article{pastor2023cross,
  title={Cross-Corpus Training Strategy for Speech Emotion Recognition Using Self-Supervised Representations},
  author={Pastor, Miguel A and Ribas, Dayana and Ortega, Alfonso and Miguel, Antonio and Lleida, Eduardo},
  journal={Applied Sciences},
  volume={13},
  number={16},
  pages={9062},
  year={2023},
  publisher={MDPI}
}
\end{small}

\end{document}